\documentclass[preprint,preprintnumbers,amsmath,amssymb]{revtex4}
\usepackage{graphicx}
\usepackage{dcolumn}
\usepackage{bm}
\usepackage{graphicx}
\usepackage{dcolumn}
\usepackage{bm}
\usepackage{rotating}
\usepackage{supertabular}
\usepackage{amssymb}
\usepackage{amsmath,theorem}
\usepackage{threeparttable}
\usepackage{multirow}
\usepackage{enumerate}
\usepackage{setspace}
\begin{document}
\date{\today}
  \newcommand{\mtr}[1]{\mbox{\boldmath${#1}$}}
  \newcommand{\bra}[1]{\mbox{$ \langle{#1}\vert$}}
  \newcommand{\ket}[1]{\mbox{$\vert{#1}\rangle$}}
  \newcommand{\ketbra}[1]{\mbox{$\vert{#1}\rangle\langle{#1}\vert$}}
  \newcommand{\ovl}[2]{\mbox{$\langle{#1}\vert{#2}\rangle$}}
  \newcommand{\vct}[1]{\mbox{$\mathbf{#1}$}}
  \newcommand{\muffintin}{R_{\operatorname{MT}}}
  \newcommand{\rkmax}{R_{\operatorname{MT}}K_{\operatorname{max}}}
\markright{}

\title{Interplay between magnetic, electronic and vibrational effects in monolayer
$\rm Mn_3O_4$ grown on Pd(100)
}

\author{C. Franchini}
\affiliation{Faculty of Physics, Universit\"at Wien and Center
for Computational Materials Science
A-1090, Wien, Austria}

\author{J. Zabloudil, R. Podloucky}                              
\affiliation{Intitut f\"ur Physikalische Chemie, Universit\"at Wien and Center
for Computational Materials Science
 A-1090, Wien, Austria}

\author{F. Allegretti, F. Li, S. Surnev, F.P. Netzer}
\affiliation{Institute of Physics, Surface and Interface Physics, 
Karl-Franzens University Graz, A-8010 Graz, Austria}


\begin{abstract}
The surface stabilized MnO(100)-like monolayer, characterised by a 
regular c(4$\times$2) distribution of Mn vacancies, is studied by hybrid functionals 
and discussed in the light of available scanning tunneling microscopy and
high-resolution electron energy loss spectroscopy data.
We show that the use of hybrid functionals is crucial to account for the intermingled
nature of magnetic ineractions, electron localization, structural distortions and 
surface phonons. 
The proposed Pd(100) supported $\rm Mn_3O_4$ structure is excellently compatible with 
the experiments previously reported in literature.
\end{abstract}

\maketitle


\section{\label{sec:intro}Introduction}
The growth of oxide ultrathin films on metal substrates represents a fascinating
and active field of research in modern surface science, primarily because of 
the particular and novel chemical and physical properties, substantially 
different and to a certain extent richer than the corresponding bulk counterparts. 
The increasing interest stems from the numerous technological applications and
the challenging fundamental understanding that have boosted intensive 
experimental and theoretical efforts. 

In this context, manganese oxides represent a distinguished example, partly because of
the complex interplay between orbital, spin and lattice degrees of freedom
which lead to prominent phenomena such as colossal magnetoresistence,  
metal-insulator transitions and exotic magnetic behaviours\cite{mott, imada, franchini07, harrison, franchinimno},
but also from a chemical point of view considering the wide range of applications 
comprising catalysis, environmental waste treatments, production of water purifying agents 
and alkaline/dry-cell batteries\cite{post, baldi, armstrong, taillefert}.
As such, thin films of manganese oxides on metal supports are promising
candidates with potential new applications in many different areas.       

Thin films of MnO have been successfully prepared on various noble metal 
substrates, such as Ag(001)\cite{muller}, Rh(001)\cite{nishimura} and Pt(111)\cite{rizzi, widdra}.
In a series of previous studies we have provided a detailed description of 
the Mn oxide phases formed on Pd(100), focusing on both high [20-30 monolayers (ML)] 
and low (below 1 ML) coverage regimes \cite{allegretti, bayer, franchini, li}. 
In particular, we have shown that upon deposition of 20-30 ML epitaxial MnO(100)
films can be grown which can be preferentially converted either into MnO(111) 
or $\rm Mn_3O_4$(001) by appropriate tuning of temperature and oxygen pressure. 
Below 1 ML, a complex surface phase diagram was reported, 
where nine different novel Mn oxide phases have been detected which belong
to two distinct oxygen pressure regimes and are characterised by well-defined
structural and vibrational properties, as established by scanning tunnelling 
microscopy (STM), low-energy electron diffraction (LEED), high-resolution
electron energy loss spectroscopy (HREELS) and density functional theory (DFT).  

Here, by means of first principles calculations, we aim to explore the structure 
and properties of one particular $\rm Mn_xO_y$-Pd(100) interfacial phase, namely
the c(4$\times$2)-$\rm Mn_3O_4$ structure which is formed at intermediate $\rm O_2$
pressures (around 5$\times$10$^{-7}$ mbar). Interestingly, this c(4$\times$2)-$\rm Mn_3O_4$/Pd(100) 
structure appears to be very similar to the recently observed c(4$\times$2)-$\rm Ni_3O_4$ phase 
obtained upon reactive evaporation of nickel on Pd(100),
which has been identified as a compressed substoichiometric NiO(100) monolayer characterised
by a regular rhombic distribution of Ni vacancies
\cite{agnoli1, agnoli2, ferrari}. The analogy between Pd(100)-supported
$\rm Mn_3O_4$ and $\rm Ni_3O_4$ c(4$\times$2) structures is likely to originate 
from the similarity between the corresponding bulk parent compounds.
MnO and NiO crystallise in the same rhombohedrally distorted fcc structure, 
display a type II antiferromagnetic spin ordering, and the electronic nature
is dominated by the partially filled localised 3$d$ states. However, a major
difference is observed in the lattice constants of MnO (a$_{\rm MnO}$=3.14 \AA) and 
NiO (a$_{\rm NiO}$=2.95 \AA) that renders the analogy between
$\rm Mn_3O_4$ and $\rm Ni_3O_4$ c(4$\times$2)/Pd(100) hardly predictable. 
Specifically, for the deposition of MnO and NiO on Pd(100) (a$_{\rm Pd}$=2.75 \AA),
this difference in the lattice constants results in a much
higher positive lattice mismatch for MnO (14 \%) than for NiO (7 \%).
Despite that, bulk-like MnO(100)
\cite{allegretti}
and NiO(100)
\cite{schoiswohl}
films has been both successfully grown on a Pd(100) substrate. 
Summing up, in spite of the different lattice mismatches, the strong similarities
of MnO and NiO seem to support the first glance analogy between the 
corresponding interfacial Pd(100) supported c(4$\times$2) phases. 

In the present work we address this issue and interpret the $\rm Mn_3O_4$ structure as a compressed
epitaxial MnO(100) monolayer with a rhombic c(4$\times$2) array of manganese
vacancies. Our study is mainly concerned with the {\em ab initio} investigation
of structural, electronic, vibrational and magnetic properties, but
a link to the experimental findings will be recalled when necessary.  

The paper is organised as follows: In Sec.~\ref{sec:comp} we describe 
the computational tools, whereas the results will be presented and discussed in Sec.~\ref{sec:res}.
Finally, in Sec.~\ref{sec:con} we draw conclusions.

\section{\label{sec:comp}Computational Details}

The results presented in this work were obtained using the Vienna \emph{ab initio} simulation package (VASP)~\cite{kresse} 
within standard (Kohn Sham theory) and generalised (hybrid) density functional theory (DFT)\cite{dft,hyb}. 
All calculations have been performed using the projector-augmented-wave (PAW) method within the generalised
gradient spin density approximation to the DFT in the Perdew-Burke-Ernzerhof parametrisation scheme\cite{pbe}.
Prompted by the satisfactory application of the HSE (Heyd-Scuseria-Ernzerhof)\cite{hse1,hse2} hybrid DFT scheme
on the physical properties of manganese oxides presented in 
a series of recent papers\cite{franchinimno, franchini07, cf3}, we have adopted the same hybrid formalism in the present work
to calculate the electronic properties and phonon frequencies of the most favourable models.

Unlike standard DFT, HSE employs an admixture of Hartree-Fock (HF) and PBE exchange in the construction
of the many body exchange (x) and correlation (c) functional:

\begin{equation}
E_{xc}^{\rm HSE03} = \frac{1}{4}E_{x}^{\rm HF,sr,\mu} +
        \frac{3}{4}E_{x}^{\rm PBE,sr,\mu} + E_{x}^{\rm PBE,lr,\mu} + E_{c}^{\rm PBE},
\end{equation}

where (sr) and (lr) refer to the short- and long-range parts of the respective exchange interactions, 
whereas $\mu$ controls the range separation, varying between 0.2 and 0.3~$\AA^{-1}$. We have used $\mu = 0.2$ \AA$^{-1}$.

The Pd(100) supported c(4$\times$2)-$\rm Mn_3O_4$ nanolayer has been modeled with a repeated slab constructed
by a four layers thick Pd(100) substrate and a single manganese deficient MnO(100) overlayer with stoichiometry $\rm Mn_3O_4$. 
In order to allow for a rhombic distribution of manganese vacancies a two-dimensional (2D) c(4$\times$2) unit cell 
has been utilised, which also permitted to test different magnetic configurations.
Finally, particular care has been devoted to the interface stacking between the $\rm Mn_3O_4$ overlayer and the metal substrate.
Overall we have tested three models for the vertical set registry, which will be discussed in more details in the next session. 

As regards the structural relaxation, the two bottommost Pd layers were kept fixed. To relax the remaining atomic positions 
(the topmost 2 Pd layers plus the $\rm Mn_3O_4$ overlayer) we used the interatomic forces calculated through the Hellmann-Feynmann 
theorem and the geometry was optimised until the change in the total energy was smaller than 10$^{-3}$ eV between two consecutive
ionic configurations. Well converged structural and electronic relaxations were reached at an energy cutoff of 300 eV and 
adopting a $\Gamma$-centred symmetry reduced 2D $6\times6$ Monkhorst-Pack~\cite{monk} $k$-point mesh
for Brillouin zone integrations.

To determine the stability of the various models we have compared their generalised adsorption energy $\gamma$ defined as:

\begin{equation}
\gamma = (E_{\rm slab} - E_{\rm Pd(100)} - n_{\rm Mn}\mu_{\rm Mn} - n_{\rm O}\mu_{\rm O})
\end{equation}

where $\rm E_{slab}$ indicates the total DFT energy of the $\rm Mn_3O_4$/Pd(100) system considered, 
$\rm E_{Pd(100)}$ refers to the clean Pd(100) DFT energy, whereas  $n_{\rm Mn/O}$ and $\mu_{\rm Mn/O}$ are the 
number of manganese ($n$=3) and oxygen ($n$=4) atoms and the corresponding reference energies.  
As discussed elsewhere\cite{cf4}, we set $\mu_{\rm Mn}$ and $\mu_{\rm O}$ to the energy of fcc bulk Mn 
($\gamma$-Mn) and half of the free $\rm O_2$ dimer energy, respectively.

Finally, phonon frequencies and eigenvectors have been evaluated by 
diagonalising the dynamical matrix of the most stable structure. 

\section{\label{sec:res}Results and Discussion}

\begin{figure}
\includegraphics[clip,width=1.0\textwidth]{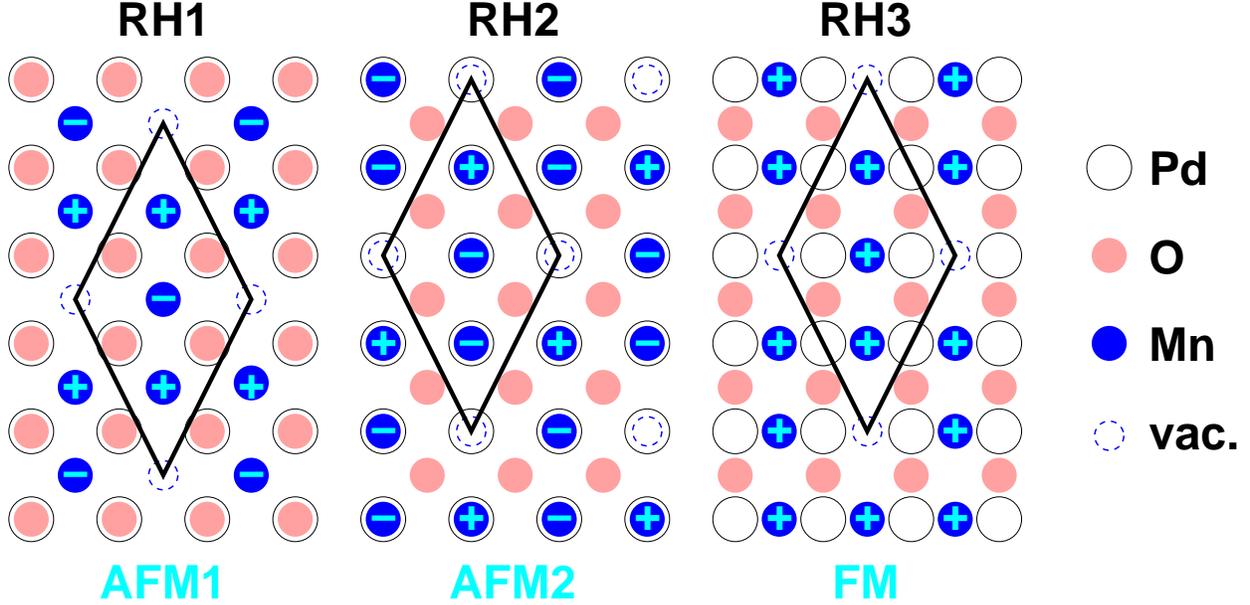}
\caption{(Colour online)
Top view of the geometrical and magnetic models explored for the Pd(100) supported c(4$\times$2) $\rm Mn_3O_4$ phase.
The models consist of a MnO(100) overlayer characterised by a rhombic distribution of Mn vacancies. Three different adsorption site registries of the 
 $\rm Mn_3O_4$ layer on the Pd atoms underneath have been considered. Models RH1 and RH2 refer to a atop O-Pd and Mn-Pd adsorption configuration, respectively, 
whereas RH3 corresponds to a bridge site adsorption of both O and Mn on Pd(100). Plus and minus signs indicate the orientation of the Mn spins
perpendicular to the surface, pointing inward and outward, respectively. 
Besides the FM arrangement two distinct AFM orderings were studied, which we have called AFM1 and AFM2.
}
\label{fig:1}
\end{figure}

The description of the structural and magnetic models considered in our study to simulate the experimentally observed c(4$\times$2) phase
are schematically depicted in Fig.\ref{fig:1}. The common building blocks consist of a Pd(100) substrate containing 4 layers on which 
a compressed MnO(100) monolayer with a rhombic distribution of Mn vacancies, resulting in an $\rm Mn_3O_4$ stoichiometry, has been placed.
As already mentioned in the introduction the lateral compression compensates for the positive mismatch between MnO and Pd.
We have therefore adopted for both substrate and interfacial layer the PBE minimised lattice constant $\rm a_{Pd}^{PBE}$ = 2.79. \AA, in good agreement
with the experimental value of 2.75 \AA. Based on the square-like termination of MnO(100) and Pd(100) three models of the interface emerge 
naturally depending on the site registry: RH1 is constructed by placing the interfacial O on top of the Pd atoms; RH2 correspond
to a top site Mn-Pd junction, whereas in the RH3 model both Mn and O are placed in the Pd(100) bridge sites. In terms of the Mn vacancies the
above models correspond to a hollow-site (RH1), top-site (RH2) and bridge-site (RH3) vacancy.
To investigate the effect of magnetism on the stability of the c(4$\times$2)-$\rm Mn_3O_4$/Pd(100) system we have computed the total energy 
of each structural model as a function of the Mn spins orientation. Besides the trivial ferromagnetic (FM) ordering, the choice of 
a c(4$\times$2) unit cell allows for the set up of two additional antiferromagnetic (AFM) orientations, AFM1 and AFM2. As displayed in 
Fig.\ref{fig:1}, AFM1 and AFM2 differ for the dissimilar spin alignment along the [100] direction which is FM and AFM, respectively.

Table \ref{tab:1} summarises the calculated relative energies for all nine possible configurations. 
The PBE values show that the RH1 is by 100-200 meV more favourable than any other geometrical setup and that
the AFM2 ordering results in the most stable magnetic order. However, the inclusion of a fraction of non-local 
Hartee-Fock exchange within the HSE approach, though favouring the RH1 geometry like PBE, reverts the magnetic relative stability 
preferring the FM configuration. Although the contradiction between PBE and HSE results is not surprising, 
in particular if related to magnetic properties of transition metal oxides, the small energy differences (tens of meV) 
and the lack of experimental magnetic analysis on this system do not permit a definitive, beyond doubt, conclusion regarding the most 
stable magnetic ordering.
In order to arrive at an indisputable and trustworthy answer a more detailed investigation of both RH1-FM and RH1-AM2 models is required, 
which discloses the interplay between the magnetic ordering and the structural, electronic and vibrational properties
for which accurate experimental findings exist. 
In the following we will therefore present in parallel the predicted physical properties of the FM and AFM2 RH1 models and
compare them with available experimental results, with the aim of determining the most favourable magnetic ground state.
 
\begin{table}
\caption{Calculated relative stability (meV/unit cell) for the magnetic models investigated.
HSE calculations have been only performed for the more favourable PBE models, 
namely RH1-FM and RH1-AFM2.
}
\vspace{0.3cm}
\begin{ruledtabular}
\begin{tabular}{lccc}
                       & FM  & AFM1 & AFM2 \\ 
\hline\hline
\\
\multicolumn{4}{c}{{PBE}} \\
\\
RH1                    &  41 &  97  &   0  \\  
RH2                    & 280 & 273  & 171  \\ 
RH3                    & 230 & 301  & 180  \\
\\
\multicolumn{4}{c}{{HSE}}\\
\\
RH1                    & 0   & -    &   22 \\  
\end{tabular}
\end{ruledtabular}
\label{tab:1}
\end{table}

\begin{figure}
\includegraphics[clip,width=1.0\textwidth]{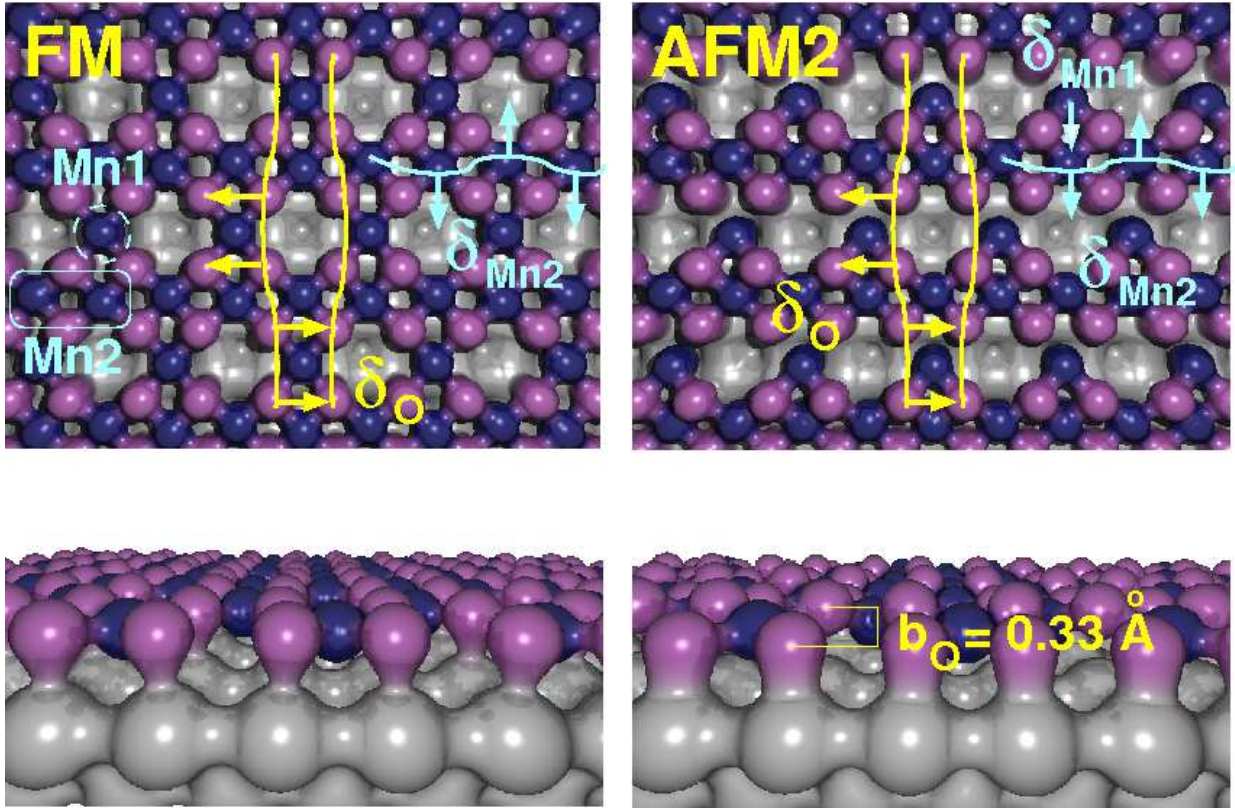}
\caption{(Colour online)
Optimised geometrical structure for the c(4$\times$2) $\rm Mn3O_4$ RH1 model in FM (left) and AFM2 (right) magnetic configuration.
The Pd(100) substrate is displayed with light gray large spheres, whereas Mn and O atoms are depicted with small dark (blue) and 
gray (pink) spheres. Two distinct Mn species are distinguishable: Mn1, sandwiched between two vacancies and two Mn2 atoms forming 
\rm the zig-zag arrangements highlighted by the horizontal lines. Atoms around the manganese vacancies experience considerable strains 
(indicated by arrows) which leads to similarly distorted FM and AFM2 structures. The only sizeable differences between the FM and AFM2
minimised geometries reside in the displacement $\delta_{\rm Mn1}$ and in the vertical buckling between oxygen atoms which are not present
in the FM model.
}
\label{fig:2}
\end{figure}

In Fig.~\ref{fig:2} we provide a graphical illustration of the optimised geometries for the RH1 FM and AFM2 models, whereas in Table \ref{tab:2}
we list the most relevant structural relaxations and magnetic moments. Although the two minimised structures appear very similar, there are important magnetically driven
differences which play a significant role, as we will discuss later on. 
First we focus our attention on the similar structural character of the explored models in terms of surface layer displacement ($\delta$), 
vertical buckling ($b$) and interface $\rm Mn_3O_4$-Pd(100) distance ($z$). Most of the relaxations take place in the $\rm Mn_3O_4$ layer
which is found to be well separated from the substrate by z$\approx$2.3 \AA, about 20\% larger than the Pd bulk interlayer distance. 
As a consequence, the Pd(100) substrate remains structurally unaffected by the growth of the $\rm Mn_3O_4$ monolayer. The presence of 
the Mn vacancies dominates the structural rearrangement and determines the appearance of two distinct Mn types: Mn1 localised between 
two nearest neighbour vacancies and Mn2 forming zig-zag Mn-Mn chains along the [100] direction. In addition, the formation of the Mn vacancy 
breaks the local Mn-O bonds and induces a planar outward relaxation of the oxygen atoms which move closer to Mn1, resulting 
in the waving long-range arrangements highlighted in Fig.~\ref{fig:2}. The lateral oxygen displacement $\delta_{\rm O}$=0.12 \AA~is found to be identical 
for both FM and AFM2, whereas the movements of the manganese atoms is notably different. Within an AFM2 ordering the lateral Mn2 displacement
($\delta_{\rm Mn2}$=0.12 \AA) turns out to be 0.04 \AA~smaller than in FM and as a consequence of that a small corrugation of 0.05 \AA~along the chain
is observed in RH1-AFM2 between adjacent antiferromagnetically coupled Mn2 atoms. Finally the Mn1 atoms in RH1-AFM2 experience a shift ($\delta_{\rm Mn1}$=0.07\AA) 
towards the nearest AFM coupled Mn2, which is responsible for the smaller width of the Mn2-Mn2 zig-zag chain in RH1-AFM2 and the quite large buckling of 
0.33 \AA~between inequivalent oxygen atoms. 
Within the FM configuration the $\delta_{\rm Mn1}$ movement is forbidden by symmetry and cannot take place, thus preventing the corrugation between the oxygen species. 

\begin{figure}
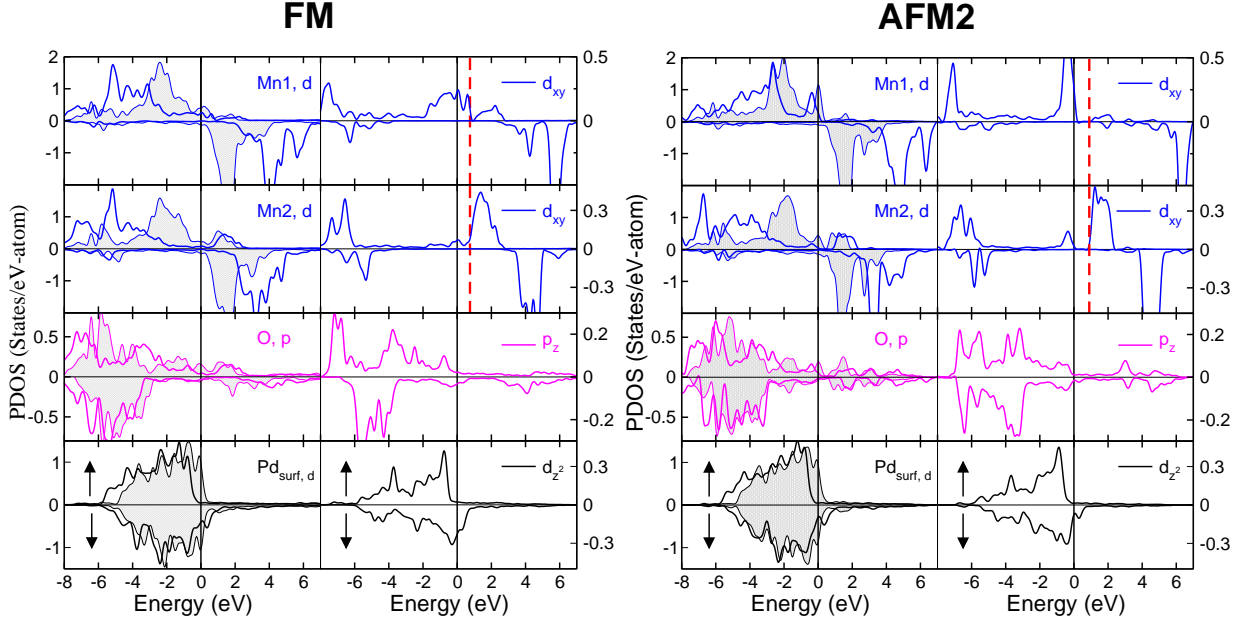

\includegraphics[clip,width=0.49\textwidth]{fig3a.eps}
\includegraphics[clip,width=0.49\textwidth]{fig3b.eps}
\caption{(Colour online)
Atom and spin resolved local density of states (LDOS) of the RH1 structure calculated within the HSE approach
in the FM and AFM2 magnetic configurations. Shadow areas (left side of panels) refer to PBE data. For  each atom, majority and minority 
spin states are plotted in different panels, as indicated by arrows.
}
\label{fig:3}
\end{figure}

As for the local magnetic moments (also reported in Table \ref{tab:2}) we do not observe any significant difference between the two 
spin arrangements. The inclusion of a fraction of HF exchange increases the local Mn moments by 10-15\% with respect to the PBE 
values and gives an induced magnetic moment of $\approx$ 0.4 $\mu_B$ on the Pd atoms directly underneath the O atoms. These HSE results are not unexpected considering
the tendency of hybrid functionals to provide a generally more localised description and a larger splitting between majority and minority states, 
as shown in the comparison between PBE and HSE Mn density of states (DOS) in Fig.~\ref{fig:3}. 

To understand the stabilisation mechanism of the RH1 phase it is instructive to make a comparison with the stoichiometric MnO monolayer.
The epitaxial growth of a perfect MnO(100) monolayer on Pd(100) is clearly unfavoured because of the large lattice mismatch between MnO and Pd.
In fact, the absence of Mn vacancies prevents the planar relaxations which give rise to the RH1 structure and lead to a very corrugated layer
with a buckling of 0.5 \AA~between O and Mn as reported in Table \ref{tab:2}. In addition, for the ideal MnO(100) system we predict a larger 
separation $z$ between adlayer and substrate which weakens substantially their mutual interaction.

\begin{table}
\caption{
Calculated optimised geometry and local magnetic moments of the $\rm Mn_3O_4$ RH1 phase compared with the ideal $\rm Mn_4O_4$ structure, 
for both FM and AFM2 magnetic orderings. The structural data refer to PBE calculations and are all given in \AA. Below the PBE magnetic moments, in
$\mu_B$, the corresponding HSE values are also listed. Symbols (schematically explained in Fig. \ref{fig:2}): $\delta$ refers to the in-plane displacements 
of Mn and O atoms; $b$'s are the vertical bucklings and $z$ indicates the average interlayer distance between the MnO layer and the Pd(100) substrate 
(also given in percent relative to the bulk Pd(100) interlayer distance).
}
\vspace{0.3cm}
\begin{ruledtabular}
\begin{tabular}{lccccccccc}
                 &$\delta_{\rm Mn1}$ & $\delta_{\rm Mn2}$ & $\delta_{\rm O}$ & $b_{\rm O-Mn}$ & $b_{\rm O-O}$ & $z$      & $m_{\rm Mn1}$ & $m_{\rm
Mn2}$ &$m_{\rm Pd}$  \\
\hline\hline
                 &               &                &              &            &         &                &           &           &          \\
                                             \multicolumn{10}{c}{{FM}}                                                             \\
                 &               &                &              &            &         &                &           &           &          \\ 
$\rm Mn_3O_4$    &        -      & 0.16           &       0.12   &    0.23    &    -    & 2.32 (+17.5\%) &      3.74 & 3.40      &   0.16   \\ 
                 &               &                &              &            &         &                &      4.20 & 3.76      &   0.42   \\ 
$\rm Mn_4O_4$    &        -      &    -           &        -     &    0.48    &    -    & 2.38 (+20.5\%) &      3.64 & 3.64      &   0.21   \\ 
                 &               &                &              &            &         &                &      4.24 & 4.24      &   0.34   \\ 
                                             \multicolumn{10}{c}{{AFM2}}                                                                 \\
                 &            &       &                  &              &  &                &                       &      &             \\
                 &            &       &                  &              &  &                &                       &      &             \\ 
$\rm Mn_3O_4$    &      0.07  & 0.12  &       0.12       &    0.23      & 0.33 & 2.36 (+19.6\%) &      3.77             & 3.26 &   0.05  \\
                 &            &       &                  &              &  &                &      4.28             & 3.70 &   0.42      \\ 
$\rm Mn_4O_4$    &        -   & -     &        -         &    0.49      & -    & 2.45 (+24.1\%) &      3.72             & 3.72 &     -   \\ 
                 &            &       &                  &              &  &                &      4.23             & 4.23 &   0.35      \\ 
\end{tabular}
\end{ruledtabular}
\label{tab:2}
\end{table}

Coming back to the nonstoichiometric adlayer, the analysis of the DOS combined with the structural properties described above allow for a comprehensive 
understanding of the bonding picture.
The hybridisation between Pd $d_{z^2}$ and O $p_z$ orbitals, particularly strong around the Fermi energy ($\rm E_F$), reflects the direct Pd-O atop site 
adsorption specific of the RH1 model. The electronic interactions within the $\rm Mn_3O_4$ layer are instead dominated by the planar bonds between
Mn1 and Mn2 $d_{xy}$ states with O $p_x$ and $p_y$ orbitals. The structural differences between Mn1 and Mn2 are clearly expressed by a dissimilar
DOS. More specifically, the Mn1-O bond, broken by the creation of the Mn vacancy, yields the formation of well localised states at $\rm E_F$ in Mn1 which
are not observed in Mn2. These localised $d$ states which characterise both FM and AFM2 magnetic alignments are responsible for the larger spin 
splitting and magnetic moment in Mn1 ($m_{\rm Mn1}\approx4.2$ $\mu_B$) as compared to Mn2 ($m_{\rm Mn2}\approx3.7$ $\mu_B$). A second important difference between 
FM and AFM2 resides in the DOS near $\rm E_F$, in the energy window between -2 and +2 eV. In the FM alignment this energy region 
shows a predominant Mn1 $d_{xy}$ spin up nature, characterised by a 3 eV broad band crossing $\rm E_F$, and only above 1 eV the Mn2 contribution
becomes 
significant. Within the AFM2 ordering the Mn1 $d_{xy}$ band near $\rm E_F$ is much more localised and displays a single peak structure which 
decays sharply at $\rm E_F$. Since the Mn2 $d$ states remain unchanged as compared to the FM situation, a gap region emerges in the range (0-0.7) eV
where the $\rm Mn_3O_4$ overlayer shows an electronic behaviour different from that predicted for the FM phase.

\begin{figure}
\includegraphics[clip,width=1.0\textwidth]{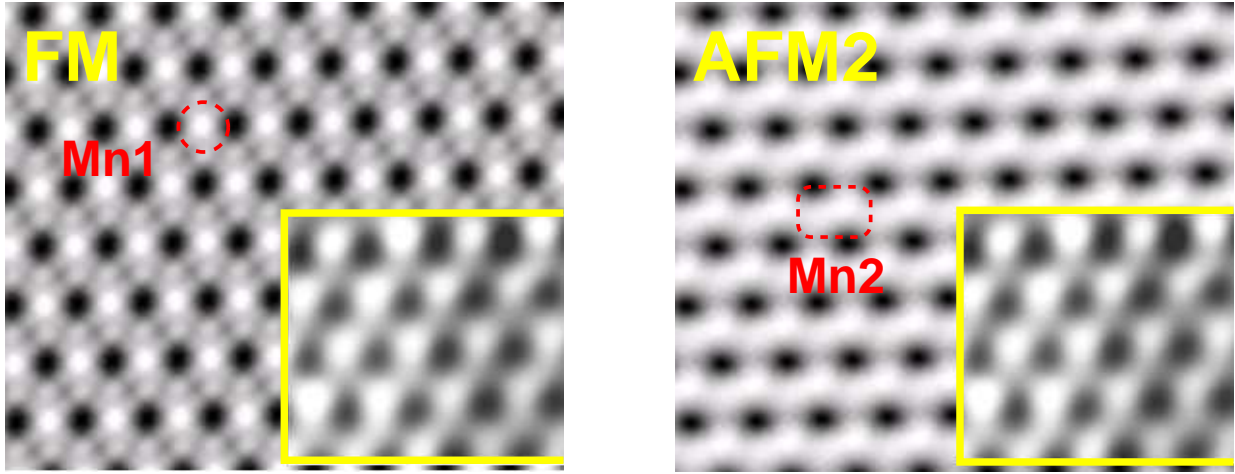}
\caption{(Colour online)
HSE RH1-FM and RH1-AFM2 simulated STM images compared with the experimental STM image (shown in the insets) taken at a sample bias U=+0.8 V and 
tunnelling current I = 0.25 nA. Oxygen atoms are not "seen", whereas manganese atoms appear as bright maxima. The detailed analysis of the 
maxima is discussed in the text.
}
\label{fig:4}
\end{figure}

At this point, comparison with the experiments is required to validate our results and to finally address the issue of which of the two modelled magnetic structures 
corresponds to the observed one. In Fig.\ref{fig:4} we compare the calculated FM and AFM2 STM images with the experimental picture.
Remarkably, the comparison unambiguously shows that the theoretical RH1-FM model resembles much better the experimental STM image as far as the contrast and the
relative brightness of the observed maxima are concerned. The measured picture is characterised by a rhombic arrangement of bright spots and dark depressions connected by weaker segments. 
These features are perfectly well reproduced by the simulated FM STM picture and can be interpreted in the following way: 
the black depressions reflect the network of Mn vacancies, the single bright spots can be assigned to the Mn1 species whereas the light segments correspond to the 
zig-zagging Mn2-Mn2 chains embedded in the regular array of oxygen atoms, which are not seen in the STM image. In the AFM2 model the Mn1 and Mn2 relative brightness
is reverted: The zig-zagging Mn2-Mn2 chains appear as the prominent feature, whereas the Mn1 species are seen as weaker unresolved connections.
Now, the following question arises naturally: why do the FM and AFM2 simulated profiles display different features? To answer this issue we note that the experimental
STM image has been taken at bias  +0.8 V, for which the FM and AFM2 phases show a very dissimilar character as discussed above and emphasised by the vertical dashed 
lines in Fig.~\ref{fig:3}. In the FM structure the STM simulation reproduces the states on top of the Mn1 $d_{xy}$ band 
centred around $\rm E_F$ weakly overlapping with Mn2-like states. In contrast, within the AFM2 configuration the only 
states contributing to the DOS near +0.8 eV come from the Mn2 $d_{xy}$ orbitals, which are therefore the ones seen 
in the AFM2 STM, in line with the interpretation given above. 

Summing up, the distinct electronic character of the FM and AFM2 phases near the experimental bias voltage is correctly recognised by HSE and is reflected in the 
simulated STM images. The FM alignment of spins is essential to reproduce the experimental STM features and the HSE approach, by reverting the PBE relative 
stability between the FM and AFM2 structures (see Table \ref{tab:1}), shows once more its predictive power and its capability of improving the description of 
transition metal compounds with respect to standard DFT. We emphasise that a standard PBE approach alone cannot provide a satisfactory and comprehensive  
understanding of this system.

\begin{table}
\caption{Predicted dipole active modes for the RH1 phase within the FM and AFM2 magnetic orderings compared to
the measured HREELS phonon peak. All data are given in meV.
}
\vspace{0.3cm}
\begin{ruledtabular}
\begin{tabular}{lccc}
                       & PBE  & HSE  & Expt. \\ 
FM                     & 36.4 & 43.3 & 43.5  \\  
AFM2                   & 35.4 & 39.6 &       \\ 
\end{tabular}
\end{ruledtabular}
\label{tab:3}
\end{table}

\begin{figure}
\includegraphics[clip,width=0.5\textwidth]{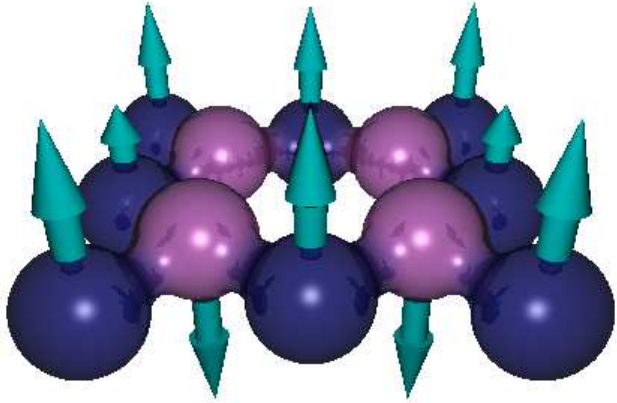}
\caption{
Schematic illustration of the atomic displacements giving rise to the calculated Mn-O vibration mode.
Mn and O atoms are sketched as black (blue) and gray (pink) spheres, respectively.  
}
\label{fig:5}
\end{figure}

Additional support for the proposed RH1-FM surface structure is provided by the comparison between experimental and theoretical vibrational properties
reported in Table \ref{tab:3}. HREELS experiments established that the phonon loss spectrum is characterised by a single peak located at 43.5 meV\cite{franchini}.
Indeed, our calculations correctly predict a single active dipole mode and demonstrate that the FM structure perfectly reproduces the experimental vibrational 
energy of 43.5 meV. Two important conclusions deserve particular enphasis: First, PBE is unable to supply a satisfactory answer, providing a much too low dipole 
active mode ($\approx$ 36 meV) and second, the AFM2 phonon peak is found to be 4 meV lower than the measured value. 
To conclude, in Fig.\ref{fig:5} we show a pictorial view of the atomic displacements associated with the vibrational peak. 
Upon diagonalisation of the dynamical matrix HSE establishes that the active dipole mode originates from the anti-phase and out-of-plane vibrations of O and 
Mn atoms, a vibrational mode typical of the ideal MnO(100) surface\cite{franchini}.

These results as well as the explanation of the STM findings are the most convincing evidence for the correctness of our computational model.

\section{\label{sec:con}Conclusions}

The formation of the two-dimensional c(4$\times$2)-$\rm Mn_3O_4$ nanolayer phase on Pd(100) has been explored by means of 
first principles PBE and HSE calculations. Combining structural, electronic, magnetic and vibrational analysis we have 
demonstrated that the preferred model which correctly describes the available experimental information entails the 
formation of rhombically distributed Mn vacancies in a compressed MnO(100) adlayer, preferentially adsorbed with an O-Pd 
atop site registry on the Pd(100) substrate. We found that hybrid functional methods such as HSE are essential to capture 
and explain all relevant properties, which would be otherwise only loosely understood within a standard DFT approach. 
In particular, HSE reverts the relative stability of the two most stable magnetic structures considered and improves
the computed vibrational energies over PBE. 
The role of the magnetic interactions is crucial to interpret the STM and HREELS experiments. 
The AFM2 ordering describes incorrectly the DOS of the $\rm Mn_3O_4$ adlayer, reverts the brightness of the observed 
STM maxima and decreases significantly the phonon dipole active mode, in contrast with the FM results which are 
fully consistent with the experimental observations.

Our study underlines the complex interplay between orbital, spin and lattice degrees of freedom occurring in 2D transition 
metal oxide nanolayers and shows that the post-DFT approach we have used represents a qualitative and quantitative step 
forward in surface science modelling.

\section{\label{sec:akn}Acknowledgments}
This work has been supported by the Austrian \emph{Fonds zur F\"orderung der wissenschaftlichen Forschung}
within the Joint Research Program S90 and by the 7$^{th}$ Framework Programme of the European Community (SEPON,
Search for Emergent Phenomena in Oxide Nanostructures).


\newpage
\begin{thebibliography}{10}

\bibitem{mott}
N. F. Mott,
Rev. Mod. Phys. {\bf 40}, 677 (1968).

\bibitem{imada}
M. Imada, A. Fujimori, Y. Tokura,
Rev. Mod. Phys. {\bf 70}, 1041 (1998).

\bibitem{franchini07}
C. Franchini, R. Podloucky, J. Paier, M. Marsman, and G. Kresse,
Phys. Rev. B {\bf 75}, 195128 (2007).

\bibitem{harrison}
W. A. Harrison,
Phys. Rev. B {\bf 77}, 245103 (2008).

\bibitem{franchinimno}
C. Franchini, V. Bayer, R. Podloucky, J. Paier, and G. Kresse, 
Phys. Rev. B {\bf 72}, 045132 (2005).

\bibitem{post}
J. E. Post, 
PNAS {\bf 96}, 3447 (1999).

\bibitem{baldi}
M. Baldi, E. Finocchio, C. Pistarino, and G. Busca,
Appl. Catal. A {\bf 173}, 61 (1998).

\bibitem{armstrong}
A. R. Armstrong and P. G. Bruce, 
Nature {\bf 381}, 499 (1996).

\bibitem{taillefert}
M. Taillefert, B. J. MacGregor, J. F. Gaillard, C. P. Lienemann, D. Perret, and D. A. Stahl, 
Environ. Sci. Technol. {\bf 36}, 468 (2002)

\bibitem{muller}
{F. M\"uller, R. de Masi, D. Reinicke, P. Steiner, S. H\"ufner, and K. Stowe,}
\newblock  Surf. Sci. {\bf 520}, 158 (2002).

\bibitem{nishimura}
{H. Nishimura, T. Tashiro, T. Fujitani, and J. Nakamura,}
\newblock  J. Vac. Sci. Technol. A {\bf 18}, 1460 (2000).

\bibitem{rizzi}
{G. A. Rizzi, R. Zanoni, S. Di Siro, L. Perriello, and G. Granozzi,}
\newblock  Surf. Sci. {\bf 462}, 187 (2000).

\bibitem{widdra}
{Ch. Hagendorf, S. Sachert, B. Bochmann, K. Kostov, and W. Widdra,}
\newblock  Phys. Rev. B {\bf 77}, 075406 (2008).

\bibitem{allegretti}
{F. Allegretti, C. Franchini, V. Bayer, M. Leitner, G. Parteder, B. Xu, A. Fleming, M. G. Ramsey, R. Podloucky, S. Surnev, and F. Netzer,}
\newblock  Phys. Rev. B {\bf 75}, 224120 (2007).

\bibitem{bayer}
{V. Bayer, R. Podloucky, C. Franchini, F. Allegretti, B. Xu, G. Parteder, M. G. Ramsey, S. Surnev, and F. Netzer,}
\newblock  Phys. Rev. B {\bf 76}, 165428 (2007).

\bibitem{franchini}
{C. Franchini, R. Podloucky, F. Allegretti, F. Li, G. Parteder, S. Surnev, and F. Netzer,}
\newblock  Phys. Rev. B {\bf 79}, 035420 (2009).

\bibitem{li}
{F. Li, G. Parteder, F. Allegretti, C. Franchini, R. Podloucky, S. Surnev, and F. P. Netzer,}
\newblock J. Phys.: Condens. Matter, accepted (2009).

\bibitem{agnoli1}
{S. Agnoli, M. Sambi, G. Granozzi, J. Schoiswohl, S. Surnev, F. P. Netzer, M. Ferrero, A. M. Ferrari, and C. Pisani,}
\newblock J. Phys. Chem. B {\bf 109}, 17197 (2004).

\bibitem{agnoli2}
{S. Agnoli, M. Sambi, G. Granozzi, A. Atrei, M. Caffio, and G. Rovida,}
\newblock Surf. Sci. {\bf 576}, 1 (2005).

\bibitem{ferrari}
{A. M. Ferrari, M. Ferrero, and C. Pisani,}
\newblock J. Phys. Chem. B {\bf 110}, 7918 (2006).

\bibitem{schoiswohl}
{J. Schoiswohl, W. Zheng, S. Surnev, M. G. Ramsey, G. Granozzi, S. Agnoli, and F. Netzer,}
\newblock Surf. Sci. {\bf 600}, 1099 (2006).

\bibitem{kresse}
{G. Kresse and J. Furthm\"uller,}
\newblock  Comput. Mater. Sci. {\bf 6}, 15 (1996);
Phys. Rev. B {\bf 54} 11169 (1996).

\bibitem{dft}
{W.~Kohn and L.~J.~Sham,}
\newblock  Phys. Rev. {\bf 140}, A1133 (1965).

\bibitem{hyb}
{A.~D.~Becke,}
\newblock J. Chem. Phys. {\bf 98}, 1372 (1993).

\bibitem{paw1}
{G.~Kresse and D.~Joubert,}
\newblock Phys. Rev. B {\bf 59}, 1758 (1999).

\bibitem{paw2}
{P.~E. Bl\"ochl,}
\newblock Phys. Rev. B {\bf 50}, 17953 (1994). 

\bibitem{pbe}
{J.~P.~Perdew, K.~Burke and M.~Ernzerhof,}
\newblock Phys. Rev. Lett. {\bf 77}, 3865 (1966).

\bibitem{hse1}
{J.~Heyd, E.~Scuseria and M.~Ernzerhof,} 
\newblock J. Chem. Phys. {\bf 118}, 8207 (2003). 

\bibitem{hse2}
{J.~Paier, R.~Hirschl, M.~Marsman and G.~Kresse,}
\newblock J. Chem. Phys. {\bf 122}, 234102 (2005). 

\bibitem{cf3}
{V.~Bayer, C.~Franchini and  R.~Podloucky,}
\newblock Phys. Rev. B {\bf 75}, 035404 (2006).  

\bibitem{monk}
{H. Monkhorst and J.~D. Pack,}
\newblock Phys. Rev. B {\bf 13}, 5188 (1976).

\bibitem{cf4}
{C.~Franchini, V.~Bayer, R.~Podloucky, G.~Parteder, S.~Surnev, and F.~Netzer,}
\newblock Phys. Rev. B {\bf 73}, 234102 (2005).

\end{thebibliography}
\end{document}